\documentclass[prd,twocolumn,showpacs,floatfix,amsmath,amssymb,floatfix]{revtex4}
\usepackage{graphicx,color,dcolumn,booktabs,bm}
\usepackage{longtable,lscape}
\usepackage{txfonts}
\usepackage{overpic}
\usepackage{amssymb}
\usepackage{indentfirst}
\usepackage{feynmf}   
\usepackage{slashed}  
\usepackage{cases}
\usepackage{color}
\usepackage{multirow}
\usepackage{graphicx,color,dcolumn,booktabs,bm}

\graphicspath{{Figures/}} %

\begin{document}

\title{Newly observed $B(5970)$ and the predictions of its spin and strange partners}
\author{Hao Xu$^{1,2}$}\email{xuh2013@lzu.edu.cn}
\author{Xiang Liu$^{1,2}$\footnote{Corresponding author}}\email{xiangliu@lzu.edu.cn}
\author{Takayuki Matsuki$^3$}
\email{matsuki@tokyo-kasei.ac.jp}
\affiliation{$^1$Research Center for Hadron and CSR Physics,
Lanzhou University $\&$ Institute of Modern Physics of CAS,
Lanzhou 730000, China\\
$^2$School of Physical Science and Technology, Lanzhou University,
Lanzhou 730000, China\\
$^3$Tokyo Kasei University, 1-18-1 Kaga, Itabashi, Tokyo 173-8602,
Japan}

\begin{abstract}
Using the effective Lagrangian approach, we study the decay behavior of the newly observed $B(5970)$ meson, which is assigned as the $2^3S_1$ state in the $B$ meson family.
A more important prediction is the detailed information of the partial and total decay widths of $B(2^1S_0)$, $B_s(2^1S_0)$ and $B_s(2^3S_1)$ as the spin and strange partners of $B(5970)$, which can be tested in future experimental searches for these missing states.

\end{abstract}

 \maketitle


The experimental status of the present $B$ meson family is very similar to those of charmed and charmed-strange meson families in 2003. In the past decade, more and more charmed and charmed-strange mesons were observed, which have stimulated our extensive discussions (see Ref. \cite{Liu:2010zb} for a concise review). As shown in the particle data group (PDG) \cite{Beringer:1900zz}, the $S$-wave $B$ and $B_s$ mesons were well established. In the past   eighteen years, some candidates for the $P$-wave $B$ and $B_s$ mesons were reported in Refs. \cite{Abreu:1994hj,Akers:1994fz,Buskulic:1995mt,Barate:1998cq,Abazov:2007vq,Aaltonen:2008aa,Aaltonen:2007ah,Abazov:2007af,Aaij:2012uva}, where we notice that CDF, D0 and LHCb had made large contributions to these observations. However, the radial excitations of the $B$ meson family have been absent untill
$B(5970)^{+/0}$ \cite{Aaltonen:2013atp} is observed.

The CDF Collaboration has very recently announced the evidence of a new resonance $B(5970)^{+/0}$ in the $B^{0}\pi^{+}/B^{+}\pi^{-}$  invariant mass spectrum \cite{Aaltonen:2013atp}. The information of its mass and width gives
 \begin{eqnarray*}
M_{B(5970)^0}/\Gamma_{B(5970)^0}&=&5978\pm5\pm12/70\pm18\pm31\,\,{\rm MeV}, \\
M_{B(5970)^+}/\Gamma_{B(5970)^+}&=&5961\pm5\pm12/60\pm20\pm40\,\,{\rm MeV},
\end{eqnarray*}
which correspond to neutral and charged $B(5970)$, respectively. The observation of $B(5970)$ enriches the spectrum of the $B$ meson family.

There are many theoretical studies of bottom and bottom-strange mesons before observing $B(5970)$, where the masses and decay behaviors of some $B$ and $B_s$ mesons are studied by
using different models \cite{Godfrey:1985xj,Orsland:1998de,Falk:1995th,Eichten:1993ub,Ebert:1997nk,Di Pierro:2001uu,Matsuki:2007zza,Matsuki:2006rz,Zhong:2008kd,Luo:2009wu,Ebert:2009ua,Colangelo:2012xi}. A recent paper of Ref. \cite{Sun:2014wea} performed a systematical study of the mass spectrum and strong decay of $B$ mesons and $B_s$ mesons, which is inspired by the newly observed $B(5970)$, where the relativistic quark model \cite{Godfrey:1985xj} and the quark pair creation (QPC) model are adopted in the calculation. See also Ref. \cite{Matsuki:2011xp} on the relativistic chiral particle decay of heavy-light mesons.

A further study of $B(5970)$ by other approaches is still an interesting research topic because it not only tests
the former theoretical study of $B(5970)$  but also helps us learn the model dependence of the corresponding results. Finally, this study can deepen our understanding on the properties of $B(5970)$. Considering the above reasons, we investigate the Okubo-Zweig-Iizuka (OZI) allowed two-body strong decays of $B(5970)$ by the effective Lagrangian approach in this paper, where $B(5970)$ is regarded as the first radial excitation of $B^*(5325)$. This assignment is due to the fact that the process $B(5970)\to B\pi$ is reported \cite{Aaltonen:2013atp} and the mass value of $B(5970)$ \cite{Aaltonen:2013atp} is very close to that of $B(2^3S_1)$ meson given in Refs. \cite{Godfrey:1985xj,Di Pierro:2001uu,Orsland:1998de,Matsuki:2006rz,Ebert:2009ua,Sun:2014wea}.
Calculating the ratios of the partial decay widths of $B(5970)$ and combining them with the experimental total width of $B(5970)$, we can give the detailed information of the partial decay widths of $B(5970)$, which is valuable for further experimental study on $B(5970)$. Besides the observed $B(5970)$, there exists its spin partner $B(2^1S_0)$, which is still missing in experiment. Thus, its identification in future experiment is an important task. In this work we predict the decay behavior of $B(2^1S_0)$. In addition, we also focus on the decay behavior of the strange partners of $B(5970)$ and $B(2^1S_0)$, which give crucial information to future experimental search for these two bottom-strange mesons.
In the following, we will present the details of how to estimate the decay widths of these states.

\if This article is organized as follows. In Sec. \ref{sec2},  we introduce the effective Lagrangian, which is constructed by both chiral symmetry and heavy quark symmetry.  Later, with the effective Lagrangian, we illustrate the calculation of the two-body OZI allowed strong decays of $B(5970)$, its spin partner $B(2^1S_0)$ and their strange partners. The last section is devoted to a short summary.\fi


In this work, we adopt the effective Lagrangian approach to study the OZI allowed strong decays of $B(5970)$, $B(2^1S_0)$ and their strange partners.

For the heavy-light meson containing one heavy quark $Q=c$ or $b$, there exist heavy quark limit and chiral symmetry.
Thus, the heavy quark effective theory (HQET) is applied to well describe the properties of heavy-light mesons (see Ref. \cite{Neubert:1993mb} for a detailed review). Similar to a hydrogen atom, the heavy-light meson is treated as a system, where the heavy quark is a static color source in the infinite heavy quark mass limit and  the remaining light quark interacts with this color source. In the heavy quark limit, the spin of the heavy quark is separated from the angular momentum of the light degrees of freedom, which is conserved separately. For a heavy-light meson system, there exists so called the heavy quark spin-flavor symmetry.

This symmetry makes the heavy-light mesons depend only on light degrees of freedom, which allows us to group the mesons into different doublets.
Denoting a heavy-light meson as $Q\overline{q}$, the total angular momentum can be expressed as ${\vec J}={\vec s}_Q+{\vec j}_\ell$. Here, ${\vec s}_Q$ is the heavy quark spin while ${\vec j}_\ell$ is the sum of the light quark spin and orbital angular momentum, i.e., $\vec j_\ell=\vec s_q+ \vec \ell$, which in fact corresponds to the angular momentum of light degrees of freedom. Consequently, two $S$-wave states with $\ell=0$ have $j_\ell^P=1/2^-$, which can be assigned to a doublet with $J^P=(0^-,1^-)$. The P-wave states with $\ell=1$ form two doublets with $J^P=(0^+,1^+)$ and $J^P=(1^+,2^+)$, which correspond to $j_\ell^P=1/2^+$ and $j_\ell^P=3/2^+$, respectively. For $D$-wave states with $\ell=2$, we find two independent results of $j_\ell^P$, i.e., $j_\ell^P=3/2^-$ and $j_\ell^P=5/2^-$, where the corresponding doublets are $J^P=(1^-,2^-)$ and $J^P=(2^-,3^-)$. The same classifications of heavy-light mesons can be naturally applied to the corresponding radial excitations.

By adopting the above approach, it is convenient to combine two meson fields into one doublet, where this effective doublet field is a $4 \times 4$ matrix \cite{Georgi:1990um}.
Let us denote $H_a$, $S_a$, and $T_a$ as the doublet meson fields corresponding to
$j_\ell^P=1/2^-$, $1/2^+$ and $3/2^+$ \cite{Falk:1992cx}, respectively, i.e.,
\begin{eqnarray*}
H_a &=&
\frac{1+{\rlap{v}/}}{2}[P_{a\mu}^*\gamma^\mu-P_a\gamma_5],
 \\
S_a &=&
\frac{1+{\rlap{v}/}}{2}[P_{1a\mu}^\prime\gamma^\mu\gamma_5-P_{0a}^*],
\\
T_a^\mu &=&\frac{1+{\rlap{v}/}}{2}
\left\{ P^{*\mu\nu}_{2a} \gamma_\nu
-P_{1a\nu} \sqrt{3 \over 2} \gamma_5 \left[ g^{\mu
\nu}-{\gamma^\nu \over 3} (\gamma^\mu-v^\mu) \right] \right\} , 
\end{eqnarray*}
where $a=u,d,s$ is the flavor index and $v$ is the meson four-velocity. The field operators contain a factor $\sqrt{m_P}$ and have dimension 3/2, where $m_P$ denotes the degenerate mass of the corresponding meson doublet.

When considering the heavy-light meson decays with the emission of a light pseudoscalar meson, we also need to introduce a chiral field $ \xi=e^{i {\cal M} /
f_\pi}$, where $\cal M$ is a matrix of the octet of light pseudoscalar mesons
with $f_\pi=132$ MeV. Thus, the strong interaction of heavy quark mesons with light pseudoscalar mesons can be described by an effective Lagrangian with both chiral symmetry and spin-flavor symmetry.
In the following, we give the effective Lagrangians of decay process $A\to H+M$, where $A=H^\prime$, $S$ and $T$, and $M$ denotes a light pseudoscalar meson. $H^\prime$ is the first radial excitations of S-wave states. At the leading order of the heavy quark mass and the light meson field expansion, we can write out the interaction terms in the Lagrangian \cite{Wise:1992hn,Colangelo:2007ds}:
\begin{eqnarray}
{\cal L}_{H^\prime\to H+M} &=& \,  g \, Tr [{\bar H}_a H^\prime_b \gamma_\mu \gamma_5 {\cal
A}_{ba}^\mu ], \label{H} \\
{\cal L}_{S\to H+M} &=& \,  h \, Tr [{\bar H}_a S_b \gamma_\mu \gamma_5 {\cal
A}_{ba}^\mu ]\, + \, h.c. , \\
{\cal L}_{T\to H+M} &=&  {h^\prime \over \Lambda_\chi}Tr[{\bar H}_a T^\mu_b
(i D_\mu {\slashed{\cal A}}+i{\slashed{D}} { \cal A}_\mu)_{ba} \gamma_5
] + h.c. ,\label{T}
\end{eqnarray}
where ${\cal A}_{\mu ba}=\frac{i}{2}\left(\xi^\dagger\partial_\mu
\xi-\xi
\partial_\mu \xi^\dagger\right)_{ba}$ 
and $D_\mu$ is the covariant derivative
$D_{\mu ba}=-\delta_{ba}\partial_\mu+{\cal V}_{\mu ba}$
with
${\cal V}_{\mu b a}=\frac{1}{2}\left(\xi^\dagger\partial_\mu \xi +\xi\partial_\mu \xi^\dagger\right)_{ba}$ .
In the above expressions, $\Lambda_\chi$ is the chiral symmetry breaking scale. The coupling constants $g,h,h^\prime$ are unknown, which can be determined by the experimental data and/or theoretical calculation.
It is noticeable that the conjugate field operator $\bar H$ creates a doublet $H$ field. These effective Lagrangians are invariant under both chiral transformation of the light chiral fields and the spin-flavor transformation of the heavy quark field.


Before illustrating the calculation of the OZI allowed two-body decays of $B(5970)$ and its partners, we first list their allowed decay channels in Table \ref{decay channel}. For these allowed decay channels of $B(5970)$, $B(2^1S_0)$, $B_s(2^3S_1)$ and $B_s(2^1S_0)$, we take the predicted masses in Ref. \cite{Godfrey:1985xj} as input if these bottom and bottom-strange mesons in the corresponding decay channels are missing in experiment.

\renewcommand{\arraystretch}{1.4}
\begin{table}[htbp]\centering
\caption{The allowed decay channels of $B(5970)$, $B(2^1S_0)$, $B_s(2^3S_1)$ and $B_s(2^1S_0)$. Here, we use $\surd$ and -- to denote the allowed and forbidden decay modes, respectively. In addition, we adopt  $1P(1^+)$ and $1P^\prime(1^+)$ to distinguish two $1^+$ states with the $(0^+,1^+)$ and $(1^+,2^+)$ doublets, respectively.}\label{decay channel}
\begin{tabular}{lcc|lcc}
\toprule[1pt] Channels                 &$B(5970)$
&$B(2^1S_0)$       &Channels       &$B_s(2^1S_0)$           &$B_s(2^3S_1)$
\\  \midrule[1pt] $\pi B$                 &$\surd$
&--       &$K B$       &--           &$\surd$
\\  $\pi B^*$                 &$\surd$
&$\surd$       &$K B^*$       &$\surd$           &$\surd$
\\  $\eta B$                 &$\surd$
&--       &$\eta B_s$       &--           &$\surd$
\\  $\eta B^*$                 &$\surd$
&$\surd$       &$\eta B^*_s$       &$\surd$           &$\surd$
\\  $K B_s$                 &$\surd$
&--       &$K^* B$       &--           &--
\\  $K B^*_s$                 &$\surd$
&--       &$K^* B^*$       &--           &--
\\  $\pi B(1^3P_0)$                 &--
&$\surd$       &$K B(1^3P_0)$       &--           &--
\\   $\pi B(1^3P_2)$                 &$\surd$
&$\surd$       &$K B(1^3P_2)$       &--           &--
\\  $\pi B(1P(1^+))$                 &$\surd$
&--       &$K B(1P(1^+))$       &--           &--
\\   $\pi B(1P^\prime(1^+))$                 &$\surd$
&--       &$K B(1P^\prime(1^+))$       &--           &--
\\
\bottomrule[1pt]
\end{tabular}
\end{table}

In the following, we apply these effective Lagrangians to study the strong decay behaviors
of $B(5970)$, $B(2^1S_0)$, $B_s(2^3S_1)$ and $B_s(2^1S_0)$.


As mentioned above, we treat $B(5970)$ as the first radial excitation of $B^*(5325)$ with spin-parity $J^P=1^-$. Thus, the discussed $B(5970)$ in this work belongs to $H^\prime$ doublet with $j_\ell^P={1/2}^-$. As shown in Table \ref{decay channel}, there are nine allowed decay channels for $B(5970)$.
Since $B(5970)\to \pi B(1^3P_2) $ occurs via $D$-wave contribution and with small phase space, the partial decay width of this process should be suppressed. Although $B(5970)\to \pi B(1P(1^+)), \pi B(1P^\prime(1^+))$ are pure $S$-wave decays, there still exists a suppression factor due to their own small phase space. 
Taking into account these considerations, we consider the following decay processes: $B(5970) \to  B \pi,\, B \eta,\, B_s K,\, B^* \pi,\, B^* \eta,\, B^*_s K$. We define five ratios: 
$R_1=\frac{\Gamma(B(5970)\to B^* \pi) }{\Gamma(B(5970)\to B \pi)}$, $R_2 =\frac{\Gamma(B(5970) \to B \eta) }{ \Gamma(B(5970) \to B \pi)}$, $R_3=\frac{\Gamma(B(5970)\to B_s K) }{ \Gamma(B(5970) \to B \pi) }$, 
$R_4 = \frac{\Gamma(B(5970)\to B^* \eta) }{ \Gamma(B(5970)\to B \pi)}$, and 
$R_5=\frac{\Gamma(B(5970) \to B^*_s K) }{ \Gamma(B(5970)\to B \pi) }$
with $B \pi=B^0 \pi^+ + B^+ \pi^0$, $B^* \pi=B^{*0} \pi^+ + B^{*+} \pi^0$, $K=K^+$ for a charged $B(5970)$, and $B \pi=B^0 \pi^0 + B^+ \pi^-$, $B^* \pi=B^{*0} \pi^0 + B^{*+} \pi^-$, $K=K^0$ for a neutral $B(5970)$.

Using the effective Lagrangian shown in Eq. (\ref{H}), we deduce the general expressions for the partial widths of $B(5970)$ decays into $B_{(s)}^{(*)}$ plus a light pseudoscalar meson, i.e.,
\begin{eqnarray}
\Gamma(B(5970) \to H_P P) &=& C_P {g^2 \over 6
\pi f_\pi^2}{M_{H_P} \over M_{B(5970)}} |\vec q|^3 ,\label{width1}\\
\Gamma(B(5970) \to H_V P) &=& C_P {g^2 \over 3
\pi f_\pi^2}{M_{H_V} \over M_{B(5970)}} |\vec q|^3
\label{width2},
\end{eqnarray}
with $|\vec q|=\left[\lambda(M_{B(5970)}^2,M_P^2,M^2_{H_{P(V)}})\right]^{1/2}/2M_{B(5970)}$, where $\lambda(a,b,c)=a^2+b^2+c^2-2ab-2ac-2bc$ is the K\"{a}llen function. In Eqs. (\ref{width1})-(\ref{width2}),
$H_P/H_V$ stands for a heavy pseudoscalar/vector meson while $P$ denotes a light pseudoscalar meson. The coefficient $C_P$ depends on the concrete light pseudoscalar meson, i.e., $C_{\pi^+} =1$, $C_{K^+} =1$, $C_\eta ={2/ 3}$ or $1/6$, $C_{\pi^0} ={1 /2}$, and $C_{K^0} ={1}$.
We should notice that these five ratios are independent on the coupling constant $g$.

\renewcommand{\arraystretch}{1.5}
\begin{table}[htbp]
\caption{The calculated ratios of $B(5970)^+$ and $B(5970)^0$ and the comparison with the result in Ref.  \cite{Sun:2014wea}.}
\label{ratios table}
\begin{center}
\begin{tabular}{cccc} \toprule[1pt]
&\multicolumn{2}{c}{This work}& \multirow{2}{*}{Ref. \cite{Sun:2014wea}}\\
     & $B(5970)^+  $ & $B(5970)^0$ \\ \midrule[1pt]
  $R_1$ & $1.64\pm 0.01$ & $1.65\pm 0.01$  &$2.53$\\
  $R_2$ & $0.025\pm 0.003$ & $0.028\pm 0.003$ &$0.13$\\
  $R_3$ & $0.084\pm 0.017$ & $0.096\pm 0.016$ &$0.05$\\
  $R_4$ & $0.026\pm 0.006$ & $0.031\pm 0.006$ &$0.22$\\
  $R_5$ & $0.061\pm 0.026$ & $0.080\pm 0.027$ &$0.06$\\\bottomrule[1pt]
\end{tabular}
\end{center}
\end{table}

With the above preparation, we can estimate the values of ratios $R_i$ ($i=1,\cdots, 5$), which are listed in Table \ref{ratios table}, where the errors are from
the experimental one for the mass of $B(5970)$ \cite{Aaltonen:2013atp}.  The results in Table \ref{ratios table} reveal $B \pi$ and $B^* \pi$ are the main decay channels of $B(5970)$. In addition, we also find that its partial decay widths of other decay channels are quite small due to the small phase space. At present, $B(5970)$ was only reported in its $B\pi$ decay channel. Thus, we suggest the experimental search for $B(5970)$ via its $B^* \pi$ channel. In addition,
it is worthwhile to notice that some results shown in Table \ref{ratios table} are comparable with those given in Ref. \cite{Sun:2014wea} (see Table \ref{ratios table} for the details), where the QPC model is adopted.

Assuming that the sum of these partial decay widths of discussed five decay modes is a dominant contribution to the total width of $B(5970)$, we can give the information of the branching fractions of these five partial decays using the ratios listed in Table \ref{ratios table}, whose concrete results are given in Table \ref{branch table}.

\renewcommand{\arraystretch}{1.5}
\begin{table*}[htbp]
\caption{The obtained branching  fractions of $B(5970)^+$ and $B(5970)^0$.}
\label{branch table}
\begin{center}
\begin{tabular}{ccccccccccccc} \toprule[1pt]
$B(5970)^+\to$     & $B^+ \pi^0  $ & $B^0 \pi^+  $ & $B^{*+} \pi^0 $ & $B^{*0} \pi ^+ $ &$B^+ \eta$ &$B_s K^+$ &$B^{*+} \eta$ &$B^*_s K^+$ \\ \midrule[1pt]
Branching ratio   & $(11.8\pm 0.3)\% $ & $(23.5\pm 0.5)\%$ & $(19.4\pm 0.3)\%$ & $(38.5\pm 0.6)\%$ & $(0.9\pm 0.1)\%$ & $(3.0\pm 0.5)\%$ & $(0.9\pm 0.2)\%$ & $(2.1\pm 0.9)\%$  \\ \midrule[1pt]
 $B(5970)^0\to$     & $B^0 \pi^0  $ & $B^+ \pi^-  $ & $B^{*0} \pi^0 $ & $B^{*+} \pi ^- $ &$B^0 \eta$ &$B_s K^0$ &$B^{*0} \eta$ &$B^*_s K^0$ \\ \midrule[1pt]
  Branching ratio& $(11.6\pm 0.3)\% $ & $(23.1\pm 0.5)\%$ & $(19.1\pm 0.3)\%$ & $(38.1+\pm 0.6)\%$ & $(1.0\pm 0.1)\%$ & $(3.3\pm 0.5)\%$ & $(1.1\pm 0.2)\%$ & $(2.8\pm 0.9)\%$  \\ \bottomrule[1pt]
\end{tabular}
\end{center}
\end{table*}

By combining the experimental width of $B(5970)$ with the obtained branching ratios, the universal coupling constant $g$ in Eqs. (\ref{width1})-(\ref{width2}) can be extracted as
\begin{eqnarray}
g &=& 0.148\pm 0.009.\label{g}
\end{eqnarray}
We notice that there were discussions of the similar coupling constants in Refs. \cite{Colangelo:2007ds,Colangelo:1994es}. Our $g$ value is comparable with the results given in these references.
The value of $g$ is important to estimate  the decay behavior of $B(2^1S_0)$, $B_s(2^1S_0)$ and $B_s(2^3S_1)$. In the following subsections, we will further predict the total and partial decay widths of these $B/B_s$ mesons.

We need to emphasize that in this work we extract the $g$ value given by Eq. (\ref{g}) by using the leading-order formula \cite{Falk:1995th}. As indicated in Ref. \cite{Fajfer:2006hi}, the one-loop chiral correction can be important to obtain the bare $g$. 
{A pion, emitted in the transition $B(5970)
\to B\pi$, has energy of about 550 MeV, which assures that
corrections ${\cal O}(p_\pi/\Lambda)$ to the presented decay rates,
both from higher-order operators (which would introduce new terms to
Eqs. (\ref{H})-(\ref{T}), suppressed by powers of $1/\Lambda$, where $\Lambda$ is a
chiral scale) and from the chiral loops would be very large.} These effects can be considered in future work if more information on $B$ meson strong decays can be provided, which is also an intriguing research topic.


It is interesting in studying $B(2^1S_0)$, which is the spin partner of $B(5970)$. Until now, $B(2^1S_0)$ is still absent in experiment. Thus, the prediction of its partial and total decay widths can provide abundant information to future experimental study of $B(2^1S_0)$. The OZI allowed strong decay modes of $B(2^1S_0)$
are listed in Table \ref{decay channel}. The mass of $B(2^1S_0)$ is taken as $5890\pm30$ Mev which covers the theoretical predictions of its mass in Refs. \cite{Matsuki:2006rz,Ebert:2009ua,Di Pierro:2001uu,Godfrey:1985xj,Sun:2014wea}.

Considering the small phase space of $B(2^1S_0)\to\pi B(1^3P_0),\, \pi B(1^3P_2)$, we do not include these two decays in our study. The relevant decay channels in this work are $B(2^1S_0)$ decaying into $B^{*0}\pi ^+$, $B^{*+}\pi ^0$ and $B^{*+}\eta$. Thus, we write out the general expression for these three decays
\begin{eqnarray}
\Gamma(B(2^1S_0)\to H_V P) &=& C_P {g^2 \over 2
\pi f_\pi^2}{M_{H_V} \over M_{B(2^1S_0)}} |\vec q^\prime|^3
\label{width3},
\end{eqnarray}
where $|\vec q^\prime|$ has the same form as that in Eqs. (\ref{width1})-(\ref{width2}).

Treating the sum of the partial decay widths of discussed three decay channels as the total width of $B(2^1S_0)$, we obtain the information of branching ratios of $B(2^1S_0)^+\to B^{*0}\pi ^+ ,\,B^{*+}\pi ^0,\,B^{*+}\eta$ which are $(66.3 \pm 0.2)\%$, $(33.4 \pm  0.1)\%$, $(0.3 \pm 0.3)\%$, respectively. The numerical result indicates that $B^*\pi$ is the dominant decay channel of $B(2^1S_0)$. Hence, it is suitable to search for the missing $B(2^1S_0)$ in the $B^*\pi$ decay mode.


Besides giving the partial decay widths of $B(2^1S_0)$, we can also get the total width of $B(2^1S_0)$ by adopting the $g$ value in Eq. (\ref{g}) as the input, i.e., $\Gamma(B(2^1S_0)^+)=40\pm11$ MeV,
which shows that $B(2^1S_0)$ has a narrow width. This prediction can be tested by future experiment since
it is not difficult to detect a narrow state.
In addition, we also compare the above total width of $B(2^1S_0)$ with the former theoretical prediction in Ref. \cite{Sun:2014wea}, where our partial and total widths are consistent with their results. {Here, we need to emphasize that the estimated total width of $B(2^1S_0)$ does not include the contribution from the $B(2^1S_0)\to B(1^3P_0)\pi,\,B(1^3P_2)\pi$ decays since these two decay modes are suppressed by the small phase phases. }


There are the strange partners of $B(5970)$ and $B(2^1S_0)$, which correspond to $B_s(2^3S_1)$ and $B_s(2^1S_0)$, respectively. If taking the masses of $B_s(2^3S_1)$ and $B_s(2^1S_0)$ to be $6000\pm30$ MeV and $5980\pm30$ MeV, respectively, which can cover former theoretical values of  the masses of $B_s(2^3S_1)$ and $B_s(2^1S_0)$ predicted in Refs. \cite{Ebert:2009ua,Di Pierro:2001uu,Matsuki:2006rz,Godfrey:1985xj,Sun:2014wea}, the OZI allowed two-body strong decays are  given in Table \ref{decay channel}.

For $B_s(2^1S_0)$, we consider the following decay channels $B_s(2^1S_0) \to B^* K,\,B^*_s \eta$. The partial and total decay widths can be obtained by using Eq. (\ref{width3}).
For $B_s(2^3S_1)$,  the allowed decay channels are $B K,\,B^*K,\,B_s \eta,\,B^*_s \eta$, where the corresponding partial and total widths are calculated by using Eq. (\ref{width1}). Since we can adopt the same approach as that used in studying $B(5970)$ and $B(2^1S_0)$, we do not describe the details here. The obtained results are shown in Table \ref{width table}. The results in Table \ref{width table} indicate:
\begin{enumerate}
\item $B^*K$ is the dominant decay channel for $B_s(2^1S_0)$, while $B K$, $B^*K$ are the dominant decay channels for $B_s(2^3S_1)$.
\item Both $B_s(2^1S_0)$ and $B_s(2^3S_1)$ are narrow states. Hence, experimental searches for these two missing states are possible by searching for their dominant decay modes.
\end{enumerate}

It is also apparent that the total decay width of $B_s(2^1S_0)$ or $B_s(2^3S_1)$ is smaller than that of $B(2^1S_0)$ or $B(5970)$, which can be well understood. This is because as for the decays of $B_s(2^1S_0)$ or $B_s(2^3S_1)$, the corresponding phase space is smaller than that of $B(2^1S_0)$ or $B(5970)$. We take the same $g$ value when discussing these bottom and bottom-strange mesons.

\begin{table*}[htpb]
\caption{The partial and total decay widths for $B_s(2^1S_0)$ and $B_s(2^3S_1)$ (in units of MeV).\label{r1}}
\label{width table}
\begin{center}
\begin{tabular}{ccccccccccc } \toprule[1pt]
     & $B^0 \bar K^0  $ & $B^+ K^-  $ & $B^{*0} \bar K^0 $ & $B^{*+} K^- $ &$B_s \eta  $ &$B^*_s \eta $ &Total \\ \midrule[1pt]
  $B_s(2^1S_0)$ & -- & -- & $12.30\pm 5.07$ & $12.68\pm5.15$ &-- & $0.72\pm0.72$ & $25.70\pm 10.95$  \\
  $B_s(2^3S_1)$ & $7.17\pm 2.41$ & $7.33\pm 2.44$ & $9.91\pm 3.81$ & $10.18 \pm3.86$ & $1.22\pm 0.74$ & $0.85\pm 0.81$ & $36.66\pm 14.06$  \\
  \bottomrule[1pt]
\end{tabular}
\end{center}
\end{table*}

{We need to emphasize that the error estimates in Tables \ref{ratios table}-\ref{r1} are not taking into account possible very large effects from chiral corrections and only deal with uncertainties on the input parameters.}


In summary, stimulated by recent observation of $B(5970)$ \cite{Aaltonen:2013atp}, we have carried out the investigation of $B(5970)$ and its spin partner $B(2^1S_0)$ using the effective Lagrangian approach. In this study, we have obtained the information of the partial decay widths of $B(5970)$. What is more important is that we have extracted the value of the universal coupling constant $g$, which has been applied to estimate the decay behavior of $B(2^1S_0)$.
The similar approach has been applied to study the strange partners of $B(5970)$ and $B(2^1S_0)$, both of which are also missing in experiments. We have obtained the partial and total decay widths of $B_s(2^1S_0)$ and $B_s(2^3S_1)$, {which can be tested in future experiment}.

The observation of $B(5970)$ has opened a window to find the radial excitations of bottom and bottom-strange mesons. It is an exciting time of searching for higher bottom and bottom-strange mesons.  We also expect more experimental progresses on  this research field, especially at LHCb, CDF, and D0.

Note added: After completing this work, a theoretical paper on $B_1(5721)$, $B_2(5747)$, $B_{s1}(5830)$, $B_{s2}(5840)$ and $B(5970)$ has appeared by using the heavy meson effective theory \cite{Wang:2014cta}, where their strong decay behaviors are given.

\vfil

\noindent {\bf Acknowledgments}: This project is supported by the National Natural Science
Foundation of China under Grants No. 11222547, No. 11175073, and No. 11035006, the Ministry of Education of China
(SRFDP under Grant No.
2012021111000), the Fok Ying Tung Education Foundation
(No. 131006).

\end{document}